# Modeling Route Choice with Real-Time Information: Comparing the Recursive and Non-Recursive Models

Xinlian Yu [a], Tien Mai [b], Jing Ding-Mastera [c], Song Gao [d], Emma Frejinger [e]

**Abstract**

Transportation systems are inherently uncertain due to disruptions such as bad weather, incident and the randomness of traveler's choices. Real-time information allows travelers to adapt to actual traffic conditions and potentially mitigate the adverse effect of uncertainty. We study the routing policy choice problems in a stochastic time-dependent (STD) network. A routing policy is defined as a decision rule applied at the end of each link that maps the realized traffic condition to the decision on the link to take next. Two types of routing policy choice models are formulated with perfect online information (POI): recursive logit model and non-recursive logit model. In the non-recursive model, a choice set of routing policies between an origin-destination (OD) pair is generated, and a probabilistic choice is modeled at the origin, while the choice of the next link at each link is a deterministic execution of the chosen routing policy. In the recursive model, the probabilistic choice of the next link is modeled at each link, following the framework of dynamic discrete choice models.

The difference between the two models results from the interplay of two sources of stochasticity, i.e., nature's probability and choice probability. The two models are equivalent when either source of stochasticity is removed, that is, in a deterministic network (as shown in Fosgerau et al., 2013) or with deterministic choice. We use an illustrative example to explore the difference between the two models when both sources of stochasticity exist, and find that when a route has state-wise stochastic dominance over the other, the recursive model predicts more extreme choice probabilities. The relation can go either way when the two routes are non-dominated.

We further compare the two models in terms of computational efficiency in estimation and prediction, and flexibility in systematic utility specification and modeling correlation.

[a] xinlianyu@umass.edu, [b] maitien86@gmail.com, [c] ding@ umass.edu, [d] sgao@ umass.edu,
[e] emma.frejinger@umontreal.ca

# 1. Introduction

Transportation systems are inherently uncertain due to disruptions such as bad weather and incidents, and the randomness of traveler's choices. Real-time information allows travelers to adapt to actual traffic conditions and potentially mitigate the adverse effect of uncertainty.

Two possible types of routing problems exist in stochastic networks: non-adaptive and adaptive. Non-adaptive routing determines a fixed path at the origin that is followed regardless of the realizations of the stochastic traffic conditions. In contrast, adaptive routing considers intermediate decision points, and a next link (or sub-path) is chosen based on collected information at each decision point. Adaptive routing is better than (or at least as good as) non-adaptive routing, since the latter can be viewed as a constrained version of the former. In this study, the term "routing policy" is used to denote the adaptive routing process. The definition of a routing policy depends on the underlying network and the information access (Gao & Chabini, 2006). Formal definitions and an example are provided in Section 2.

We study the problem of modeling routing policy choices following two different paradigms in the route choice literature. Logit models are the focus, however some results can be extended to any discrete choice models.

In conventional (non-recursive) route choice modeling, a choice set is generated for each origin-destination (OD) pair and then probabilities are assigned among the alternatives in the choice set with a multinomial logit model (see, e.g., reviews in Ramming, 2002; Frejinger, 2008). A non-recursive routing policy choice model generalizes the choice set of paths to that of routing policies (Gao, 2005; Gao & Huang, 2009) and conventional choice set generation algorithms such as link elimination and link penalties can be generalized by replacing the shortest path algorithm with optimal routing policy algorithm (Gao, 2005; Ding-Mastera et al., 2014, 2015). Note that the term "non-recursive" refers to the fact that a logit model is only applied at the origin. Each choice alternative for the logit model is a routing policy, which by definition is adaptive, and is indeed generated based on recursive equations (Gao & Chabini, 2006).

Fosgerau et al. (2013) proposed a recursive logit model for non-adaptive path choice in a deterministic network, where the path choice problem is formulated as a sequence of link choices. At the end of each link the decision maker chooses the utility-maximizing outgoing link with link utilities given by the instantaneous cost, the expected maximum utility to the destination (value function) and i.i.d. (independent and identically distributed ) extreme value error terms. It is shown that the recursive logit model is equivalent to a logit model at the origin with a choice set of all paths between the OD. The recursive model obviates the generation of choice sets, which are often difficult to have 100% coverage of the observed paths. The extension to routing policy choice entails including real-time traffic information in the state variables, and taking expectations of value functions over possible next states in the recursive equations as the network is stochastic.

We study and compare the two types of routing policy choice models. The contribution of this paper is the formulation of the recursive logit model for routine policy choice in a stochastic time-dependent network, and the comparison of the two models in predicting route choice probabilities based on theoretical analysis and illustrative examples. Insights on the sources of difference are provided and conditions under which the two models are equivalent are discussed.

Section 2 introduces the network settings and the routing policy definition. Section 3 gives the formulation of the recursive and the non-recursive logit model for routing policy choice. Section 4 compares the two formulations both in a general network and with an illustrative example. Discussions on the pros and cons in applying the models to real networks are also provided in Section 4. Section 5 concludes and discusses future directions.

**2. Network Settings and the Routing Policy Definition**

**Notations**

Network
$N$: set of nodes
$A$: set of links, and $|A| = m$
$T$: set of time periods $\{0, 1, \ldots, K-1\}$
$A(k)$: set of outgoing links from link $k$

Stochasticity
$v_r$: the $r^{th}$ support point, a vector with a dimension $K \times m$, $r = 1, 2, \ldots, R$
$P$: joint probability distribution of all link travel time random variables, $P = \{v_1, \ldots, v_R\}$
$R$: number of network support points for the joint distribution of all links at all time periods
$p_r$: probability of support point $r$, $\sum_{r=1}^{R} p_r = 1$

Information
$EV$: event collection, set of network support points compatible with the realized link travel times
$EV(t)$: set of all possible event collections at time $t$
$P_r(EV'|EV)$: conditional probability of $EV'$ from a later time given $EV$

Routing policy

$(k, t, EV)$: a state comprising of link $k$, time $t$ at the end of link $k$ and event collection $EV$

$\gamma$: $(k, t, EV) \rightarrow a$, routing policy, which is a mapping from a state to a next link $a$

$C_n(k_0, t_0, EV_0)$: choice set of routing policies conditional on the initial state $(k_0, t_0, EV_0)$ for traveler $n$

$V_{\gamma n}$: deterministic utility of routing policy $\gamma$ for traveler $n$

Utilities and choices

$u_n(a|k, t, EV)$: instantaneous utility associated with link $a$ based on $(k, t, EV)$ for traveler $n$

$\omega_n(a|k, t, EV)$: deterministic utility associated with link $a$ based on $(k, t, EV)$ for traveler $n$

$V_n^d(k, t, EV)$: the expected utility of state $(k, t, EV)$ for traveler $n$ with destination $d$

$P_n^d(a|k, t, EV)$: probability of choosing next link $a$ based on state $(k, t, EV)$ for traveler $n$ with destination $d$

$\sigma$: sequence of states, $\sigma = ((k_i, t_i, EV_i))_{i=0}^{I}$ where $(k_0, t_0, EV_0)$ is the initial state and $(k_I, t_I, EV_I)$ is the final state in the sequence

$P_n^d(k_{i+1}, t_{i+1}, EV_{i+1}|k_i, t_i, EV_i)$: probability of state $(k_{i+1}, t_{i+1}, EV_{i+1})$ conditional on $(k_i, t_i, EV_i)$

$f((k_{i+1}, t_{i+1}, EV_{i+1})|((k_i, t_i, EV_i), k_{i+1}))$: transition probability from $(k_i, t_i, EV_i)$ to $(k_{i+1}, t_{i+1}, EV_{i+1})$ by taking link $k_{i+1}$

$P_n(\sigma)$: the likelihood of observing a sequence of states $\sigma$ for traveler $n$

## 2.1. Network Settings

The network is modeled as a stochastic time-dependent (STD) network, in which link travel times are jointly distributed time-dependent random variables. Let $G = (N, A, T, P)$ denote an STD network, where $N$ is the set of nodes, $A$ the set of links with $|A|=m$, $T$ the set of time periods $\{0, 1, \ldots, K-1\}$, and $P$ the probabilistic description of link travel times. At time period $K-1$ and beyond, travel times are static and deterministic.

A support point is defined as a distinct vector of values that a discrete random vector can take. Thus a probability mass function (PMF) of a random variable (or vector) is a combination of support points and the associated probabilities. A joint probability distribution of all link travel time random variables is assumed: $P = \{v_1, \ldots, v_R\}$, where $v_i$ is a vector with a dimension $K \times m$, $r = 1, 2, \ldots, R$, and $R$ is the number of support points. The $r^{th}$ support point has a probability of $p_r$ and $\sum_{r=1}^{R} p_r = 1$.

Real-time information is assumed to include realized travel times of certain links at certain time periods. There are various kinds of real-time information access, e.g., full information, perfect online information (POI) and partial online information, see, e.g., Gao and Huang (2012) for discussions on a number of real-time information access. In this study, we formulate the problem with perfect online information which includes realized travel times on all link travel times up to the current time.

With the help of online information, the travelers become more certain about the future or the network becomes less stochastic. To model this effect of information in reducing uncertainty, the concept of the event collection, denoted as $EV$, is introduced as a subset of support points that are compatible with the realized travel times. It represents the conditional distribution of link travel times given the realization of link travel times. As more information becomes available, the size of an event collection decreases or remains the same. When an event collection becomes a singleton, the network becomes deterministic.

An illustrative example network is shown in Figure 1 and Table 1 with three nodes, four links (including a dummy link 0 with a zero travel time) and two time periods. We denote links by the numbers beside it. There are two support points, each with a probability 1/2, for the joint distribution of six (links 1, 2 and 3 at time periods 0 and 1) travel time random variables. Travel time beyond time period 1 are the same as those in time period 1 in either of the two support points. Two paths are available: link 1–link 2 (path 1) and link 1-link 3 (path 2). At time 0, there is only one possible event collection $(v_1, v_2)$, as travel times on all links are the same across the two support points at time 0. At time 1, there are two possible event collections, $v_1$ and $v_2$.

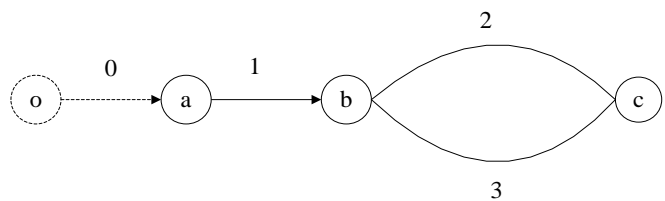

**Figure 1** A small illustrative network

**Table 1** Support points for the network ($p_1 = p_2 = \frac{1}{2}$)

| Time | Link | $v_1$ | $v_2$ |
|---|---|---|---|
| 0 | 1 | 1 | 1 |
|   | 2 | 2 | 2 |
|   | 3 | 1 | 1 |
| 1 | 1 | 1 | 2 |
|   | 2 | 3 | 2 |
|   | 3 | 2 | 2 |

## 2.2. Routing Policy

Assume the traveler make a decision at the end of each link as to which link to take next based on the current state $(k, t, EV)$, where $k$ is the current link, $t$ is the arrival time at the end of link $k$, and $EV$ is the event collection.

A routing policy $\gamma$ between an initial state and a destination is defined as a mapping from states to next links, that is, $(k, t, EV) \rightarrow a$, for all states that are reachable from the initial state and can reach the destination. Travelers who follow a routing policy make decisions en route and therefore can take different paths, depending on actual network conditions.

In the network of Figure 1, consider the following routing policy: the traveler starts with an initial state $\{0, 0, (v_1, v_2)\}$ and takes link 1 since there is no other choice; at the end of link 1 (node $b$) two states $\{1, 1, v_1\}$ or $\{1, 1, v_2\}$ are possible. At $\{1, 1, v_1\}$ the traveler takes link 2 and arrives at the destination with a final state of $\{2, 4, v_1\}$. At $\{1, 1, v_2\}$ the traveler takes link 3 and arrives at the destination with a final state of $\{3, 3, v_2\}$. This is represented intuitively in Figure 2 as a "state tree" for routing policy $\gamma_2$. Figure 2 also includes the other three routing policies, where $\gamma_1$ and $\gamma_4$ are not adaptive to states and simply fixed paths.

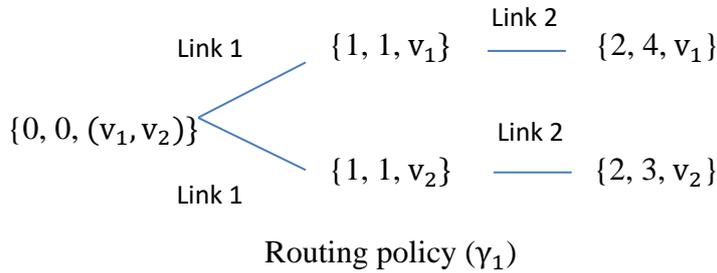

Routing policy ($\gamma_1$)

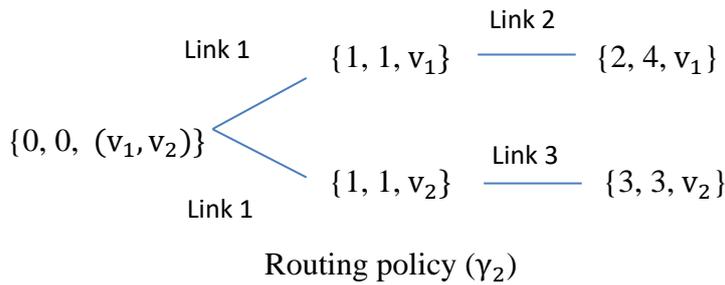

Routing policy ($\gamma_2$)

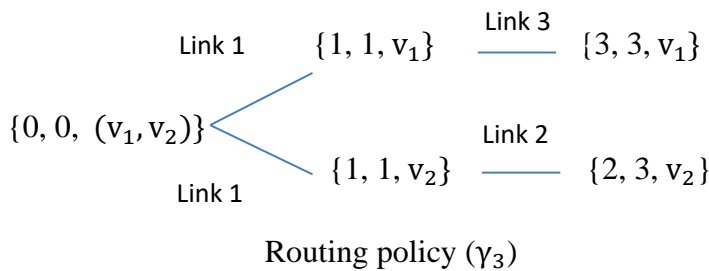

Routing policy ($\gamma_3$)

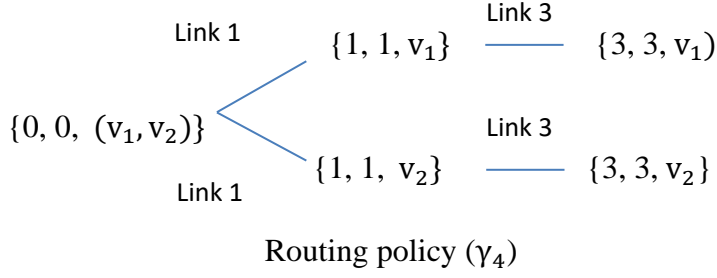

Figure 2 State trees of all possible routing policies for the network in Figure 1

## 3. Model Specification

Before formalizing the two models in this section, we briefly introduce the multinomial logit model for a set of alternatives $j \in \{1, ... J\}$. A utility $u_j$ is associated with each alternative and is the sum of a deterministic and a random component $\omega_j + \mu\varepsilon_j$, where $\varepsilon_j$ are assumed i.i.d. extreme value type 1 with zero mean and $\mu$ is scale parameter.

Let $u_{max} = max_j \mu_j$ denote the maximum utility and the expected maximum utility is $Eu_{max} = \mu ln \sum_j e^{\frac{1}{\mu}\omega_j}$. It is a general fact for additive random utility models (McFadden, 1978) that choice probabilities can be found as the gradient of $Eu_{max}$ considered as a function of the vector of deterministic utility components $\omega$ and hence

$$P_j = \frac{e^{\frac{1}{\mu}\omega_j}}{e^{\frac{1}{\mu}E\mu_{max}}} = \frac{e^{\frac{1}{\mu}\omega_j}}{\sum_{j'} e^{\frac{1}{\mu}\omega_{j'}}} \qquad (1)$$

### 3.1. The Recursive Logit Model

We formulate the routing policy choice problem as a dynamic discrete choice model where the utility maximization problem is consistent with a dynamic programming (DP) problem. The model is an extension of Fosgerau et al. (2013) to a stochastic time-dependent network.

Consider an individual travelling from an origin to a destination $d$. The traveler starts from a specific initial state and reached the next state by choosing an action $a$ (next link) from the set of outgoing links $A(k)$ from the sink node of link $k$. An instantaneous utility for traveler $n$, $u_n(a|k,t,EV) = \omega_n(a|k,t,EV) + \mu\epsilon_n(a|k,t,EV)$ is associated with each action in the choice set $A(k)$. The random terms $\epsilon_n(a|k,t,EV)$ are assumed i.i.d. Gumbel with scale parameter 1 and they are independent with everything in the model.

The traveler chooses the next link based on the current state in a stochastic process with the Markov property. At each state $(k, t, EV)$ the traveler chooses the action (link) to maximize the utility that is the sum of the instantaneous utility $u_n(a|k, t, EV)$, and the expected downstream utility, which is the expectation of the value function over all possible next states. The value function $V_n^d(k, t, EV)$ can be obtained using the recursive expression by taking the continuation of this process into account via the Bellman equation (Bellman, 1957).

$$V_n^d(k, t, EV) = E\{max_{a \in A(k)}[\omega_n(a|k, t, EV) + \Sigma_{EV' \in EV(t')} V_n^d(a, t', EV') P_r(EV'|EV) + \mu \epsilon_n(a|k, t, EV)]\} \quad (2)$$

$t'$ is the arrival time at the end of the chosen link $a$, and $t' = t + \tau(a|k, t, EV)$ where $\tau(a|k, t, EV)$ is the travel time on link $a$ conditional on current state $(k, t, EV)$. $\tau(a|k, t, EV)$ is deterministically specified based on the POI. $EV'$ is one of the possible event collection at $t'$. $P_r(EV'|EV)$ is the probability of transforming to $EV'$ from $EV$, which is computed as

$$P_r(EV'|EV) = \frac{\Sigma_{r|r \in EV' \cap EV} p_r}{\Sigma_{r|r \in EV} p_r} \quad (3)$$

where $p_r$ is the probability of support point $v_r$.

The probability of choosing link $a$ based on the current state $(k, t, EV)$ is given by the multinomial logit model

$$P_n^d(a|k, t, EV) = \frac{e^{\frac{1}{\mu}\left(\omega_n(a|k, t, EV) + \Sigma_{EV' \in EV(t')} V_n^d(a, t', EV') P_r(EV'|EV)\right)}}{\Sigma_{a' \in A(k)} e^{\frac{1}{\mu}\left(\omega_n(a'|k, t, EV) + \Sigma_{EV' \in EV(t')} V_n^d(a', t', EV') P_r(EV'|EV)\right)}} \quad (4)$$

Then the value function is the log sum.
$$V_n^d(k, t, EV)$$
$$= \begin{cases} \mu \ln \sum_a \delta(a|k, t, EV) e^{\frac{1}{\mu}\left(\omega_n(a|k, t, EV) + \Sigma_{EV' \in EV(t')} V_n^d(a, t', EV') \cdot P_r(EV'|EV)\right)}, & \forall k \in A \setminus \{d\} \\ 0, & k = d \end{cases}$$
(5)

where $\delta(a|k, t, EV) = \begin{cases} 1, & a \in A(k) \\ 0, & a \notin A(k) \end{cases}$

We transform (5) by taking the exponential and raising to the power $\frac{1}{\mu}$.

$$e^{\frac{1}{\mu} V(k,t,EV)} =$$

$$\begin{cases} \Sigma_a \sigma(a|k, t, EV) e^{\frac{1}{\mu}\left(\omega_n(a|k, t, EV) + \Sigma_{EV' \in EV(t')} V_n^d(a, t', EV') P_r(EV'|EV)\right)}, & \forall k \in A \setminus \{d\} \\ 1, & k = d \end{cases} \quad (6)$$

An example from Figure 1 and Table 1 is used to show the definition of the value function and the probability of taking next link based on current state. Assume the deterministic link utility $\omega_n\{a|k,t,EV\} = -\tau(a|k,t,EV)$. At initial state $\{0, 0, (v_1, v_2)\}$, there is only one outgoing link 1 with two possible next states $\{1,1,v_1\}$ and $\{1,1,v_2\}$. According to Equation (5), the value function of state $\{0, 0, (v_1, v_2)\}$ is

$V_n^d\{0,0,(v_1,v_2)\}$

$= \mu ln e^{\frac{1}{\mu}(\omega_n\{1|0,0,(v_1,v_2)\}+V_n^d\{1,1,v_1\}P_r((v_1)|(v_1,v_2))+V_n^d\{1,1,v_2\}P_r((v_2)|(v_1,v_2)))}$

$= \omega_n\{1|0,0,(v_1,v_2)\} + V_n^d\{1,1,v_1\}P_r((v_1)|(v_1,v_2)) + V_n^d\{1,1,v_2\}P_r((v_2)|(v_1,v_2))$

$= -1 + 1/2 V_n^d\{1,1,v_1\} + 1/2 V_n^d\{1,1,v_2\}$

where $1/2 V_n^d\{1,1,v_1\} + 1/2 V_n^d\{1,1,v_2\}$ is the expected downstream utility over the two possible next states $\{1,1,v_1\}$ and $\{1,1,v_2\}$, each with a probability of 1/2.

The probability of taking link 1 based on state $\{0, 0, (v_1, v_2)\}$ is trivially

$P\{1|0,0,(v_1,v_2)\}$

$= \dfrac{e^{\frac{1}{\mu}\left(\omega_n\{1|0,0,(v_1,v_2)\}+V_n^d\{1,1,v_1\}P_r\left((v_1)\big|(v_1,v_2)\right)+V_n^d\{1,1,v_2\}P_r\left((v_2)\big|(v_1,v_2)\right)\right)}}{e^{\frac{1}{\mu}\left(\omega_n\{1|0,0,(v_1,v_2)\}+V_n^d\{1,1,v_1\}P_r\left((v_1)\big|(v_1,v_2)\right)+V_n^d\{1,1,v_2\}P_r\left((v_2)\big|(v_1,v_2)\right)\right)}}$

$= 1$

For estimation of the model, a sequence of states is observed for each individual. Define a sequence of states $\sigma = (k_i, t_i, EV_i)_{i=0}^I$ where $(k_0, t_0, EV_0)$ is the initial state and $(k_I, t_I, EV_I)$ is the final state in the sequence.

There are four possible sequences of states in the example of Figure 1 and Table 1:

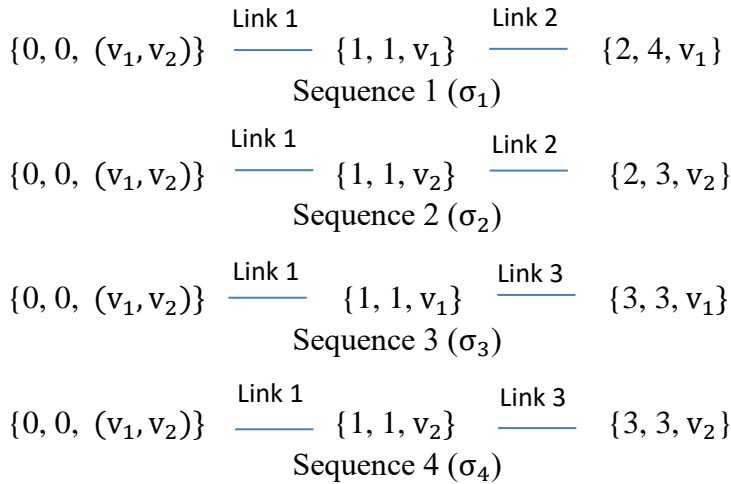

The traveler's beliefs about the next state at each intermediate state given a chosen action can be represented by a Markov transition function $f((k_{i+1}, t_{i+1}, EV_{i+1})|((k_i, t_i, EV_i), k_{i+1})$. $k_{i+1}$ is fixed given $k_{i+1}$. Travel time on $k_{i+1}$ is deterministically specified based on the POI assumption. Thus $t_{i+1}$ is also fixed given $t_i$ and $k_{i+1}$. What remains is simply the probability of ending up in $EV_{i+1}$. That is

$$f((k_{i+1}, t_{i+1}, EV_{i+1})|((k_i, t_i, EV_i), k_{i+1})) = P_r(EV_{i+1}|EV_i) \qquad (7)$$

The likelihood of observing a sequence of states is

$$P_n(\sigma) = \prod_{i=0}^{I-1} [P_n^d(k_{i+1}, t_{i+1}, EV_{i+1}|k_i, t_i, EV_i)]$$

$$= \prod_{i=0}^{I-1} P_n^d(k_{i+1}|k_i, t_i, EV_i) f((k_{i+1}, t_{i+1}, EV_{i+1})|((k_i, t_i, EV_i), k_{i+1})]$$

$$= \prod_{i=0}^{I-1} [P_n^d(k_{i+1}|k_i, t_i, EV_i) P_r(EV_{i+1}|EV_i)]$$

$$= \prod_{i=0}^{I-1} \left[\frac{e^{\frac{1}{\mu}\left(\omega_n(k_{i+1}|k_i, t_i, EV_i) + \Sigma_{EV' \in EV(t_{i+1})} V_n^d(k_{i+1}, t_{i+1}, EV') \cdot P_r(EV'|EV)\right)}}{\sum_{a \in A(k_i)} e^{\frac{1}{\mu}\left(\omega_n(a|k_i, t_i, EV_i) + \Sigma_{EV' \in EV(t')} V_n^d(a, t', EV) \cdot P_r(EV'|EV)\right)}} P_r(EV_{i+1}|EV_i)\right]$$

$$= \prod_{i=0}^{I-1} \left[\frac{e^{\frac{1}{\mu}\left(\omega(k_{i+1}|k_i, t_i, EV_i) + \Sigma_{EV' \in EV(t_{i+1})} V_n^d(k_{i+1}, t_{i+1}, EV') \cdot P_r(EV'|EV)\right)}}{e^{\frac{1}{\mu}V_n^d(k_i, t_i, EV_i)}} P_r(EV_{i+1}|EV_i)\right]$$
(8)

where $t' = t + \tau(a|k_i, t_i, EV_i)$.

Unlike the recursive logit model in a deterministic network, the enumerator in Equation (8) involves an expectation of value functions over all possible next event collections $EV' \in EV(t_{i+1})$, and not just the observed next event collection $EV_{i+1}$. As a result, the equation cannot be further simplified by canceling out value functions of adjacent links as is done in Fosgerau et al. (2013).

### 3.2. The Non-recursive Logit Model

For the non-recursive logit model, we assume that the traveler takes a routing policy from a choice set of routing policies at the origin and follows it during the trip. The choice set is generated based on a given initial state. The probability of a routing policy γ chosen by individual *n* at the origin is given by the multinomial logit model.

$$P_n(\gamma|C_n(k_0, t_0, EV_0); \beta) = \frac{e^{\frac{1}{\mu}V_{\gamma n}}}{\sum_{\gamma' \in C_n(k_0, t_0, EV_0)} e^{\frac{1}{\mu}V_{\gamma' n}}} \qquad (9)$$

$C_n(k_0, t_0, EV_0)$ is the individual specific choice set of routing policies conditional on an initial state $(k_0, t_0, EV_0)$. $\beta$ denotes the vector of parameters to be estimated. $V_{\gamma n}$ is the deterministic utility of the routing policy $\gamma$ for individual $n$.

Similar to the recursive logit model, for estimation we need to write out the likelihood of observing a sequence of states $\sigma = (k_i, t_i, EV_i)_{i=0}^I$. For non-recursive model, the adaptation en route is an execution of the chosen routing policy. A routing policy $\gamma$ contains a sequence $\sigma$, if the routing policy maps to the next observed state in $\sigma$ at every decision state, that is, $\gamma(k_i, t_i, EV_i) = k_{i+1}, i = 0, \ldots, I-1$. $\Delta(\sigma|\gamma)$ is a binary variable that equals 1 if routing policy $\gamma$ contains the observed sequence of state $\sigma$ and 0 otherwise. The probability of observing a sequence $\sigma$ given that routing policy $\gamma$ is chosen, $P(\sigma|\gamma)$, is 0 when $\Delta(\sigma|\gamma) = 0$. When $\Delta(\sigma|\gamma) = 1$, $P(\sigma|\gamma)$ is the joint probability that $(k_i, t_i, EV_i)$ transitions to $(k_{i+1}, t_{i+1}, EV_{i+1})$ given $\gamma(k_i, t_i, EV_i) = k_{i+1}, i = 0, \ldots, I-1$. According to Equation (7), it is equal to $\prod_{i=0}^{I-1} P_r(EV_{i+1}|EV_i) = P_r(EV_I|EV_0)$. Therefore

$$P(\sigma|\gamma) = \Delta(\sigma|\gamma) P_r(EV_I|EV_0) \tag{10}$$

A sequence of state can be observed by executing different routing policies with the same initial state. In the example of Figure 1 and Table 1, both routing policy $\gamma_1$ and $\gamma_2$ contain the sequence of states $\sigma_1$.

By summing over all routing policies in $C_n(k_0, t_0, EV_0)$, the probability of observing a sequence of states $\sigma$ is given as follows

$$P_n(\sigma) = \sum_{\gamma \in C_n(k_0, t_0, EV_0)} P(\gamma|C_n(k_0, t_0, EV_0); \beta) P(\sigma|\gamma) \tag{11}$$

### 3.3. Illustrative Example

Next, we continue to use the example from Figure 1 and Table 1 in Section 2 to go through the specification of the two models. Let the scale parameter for the random term $\mu$ equal 1. For the sake of simplicity the only attribute is link travel time with a parameter -1.

### 3.3.1. Recursive Model

The probability of choosing next link based on the current state is

$P\{1|0,0, (v_1, v_2)\} = 1$

$P\{2|1,1, v_1\} = \dfrac{e^{-3}}{e^{-3} + e^{-2}} = \dfrac{1}{1+e}$

$P\{2|1,1, v_2\} = \dfrac{e^{-2}}{e^{-2} + e^{-2}} = \dfrac{1}{2}$

$P\{3|1, 1, v_1\} = \frac{e^{-2}}{e^{-3}+e^{-2}} = \frac{1}{e^{-1}+1}$

$P\{3|1, 1, v_2\} = \frac{e^{-2}}{e^{-2}+e^{-2}} = \frac{1}{2}$

The likelihood of observing a sequence of states is

$P(\sigma_1) = P\{1|0,0, (v_1, v_2)\} \cdot \frac{1}{2} \cdot P\{2|1,1, v_1\} = \frac{1}{2} \cdot \frac{1}{1+e} = \frac{1}{2(1+e)} \approx 0.1345$

$P(\sigma_2) = P\{1|0,0, (v_1, v_2)\} \cdot \frac{1}{2} \cdot P\{2|1,1, v_2\} = \frac{1}{2} \cdot \frac{1}{2} = 0.25$

$P(\sigma_3) = P\{1|0,0, (v_1, v_2)\} \cdot \frac{1}{2} \cdot P\{3|1, 1, v_1\} = \frac{1}{2} \cdot \frac{1}{e^{-1}+1} = \frac{1}{2(e^{-1}+1)} \approx 0.3655$

$P(\sigma_4) = P\{1|0,0, (v_1, v_2)\} \cdot \frac{1}{2} \cdot P\{3|1, 1, v_2\} = \frac{1}{2} \cdot \frac{1}{2} = 0.25$

### 3.3.2. Non-recursive Model

The expected travel time conditional on the initial state $\{0, 0, (v_1, v_2)\}$ for each routing policy is

Routing policy 1($\gamma_1$):  $\frac{1}{2}(1+3) + \frac{1}{2}(1+2) = 3.5$

Routing policy 1($\gamma_2$):  $\frac{1}{2}(1+3) + \frac{1}{2}(1+2) = 3.5$

Routing policy 1($\gamma_3$):  $\frac{1}{2}(1+2) + \frac{1}{2}(1+2) = 3$

Routing policy 1($\gamma_4$):  $\frac{1}{2}(1+2) + \frac{1}{2}(1+2) = 3$

The probability of observing a sequence of states is

$P(\sigma_1) = \left(\frac{e^{\gamma_1}+e^{\gamma_2}}{e^{\gamma_1}+e^{\gamma_2}+e^{\gamma_3}+e^{\gamma_4}}\right) P((v_1)|EV_0) = \frac{1}{2}\left(\frac{e^{-3.5}+e^{-3.5}}{e^{-3.5}+e^{-3.5}+e^{-3}+e^{-3}}\right) = \frac{e^{-3.5}}{2(e^{-3.5}+e^{-3})} \approx 0.1888$

$P(\sigma_2) = \left(\frac{e^{\gamma_1}+e^{\gamma_3}}{e^{\gamma_1}+e^{\gamma_2}+e^{\gamma_3}+e^{\gamma_4}}\right) P((v_2)|EV_0) = \frac{1}{2}\left(\frac{e^{-3.5}+e^{-3}}{e^{-3.5}+e^{-3.5}+e^{-3}+e^{-3}}\right) = 0.25$

$P(\sigma_3) = \left(\frac{e^{\gamma_3}+e^{\gamma_4}}{e^{\gamma_1}+e^{\gamma_2}+e^{\gamma_3}+e^{\gamma_4}}\right) P((v_1)|EV_0) = \frac{1}{2}\left(\frac{e^{-3}+e^{-3}}{e^{-3.5}+e^{-3.5}+e^{-3}+e^{-3}}\right) = \frac{e^{-3}}{2(e^{-3.5}+e^{-3})} \approx 0.3112$

$P(\sigma_4) = \left(\frac{e^{\gamma_2}+e^{\gamma_4}}{e^{\gamma_1}+e^{\gamma_2}+e^{\gamma_3}+e^{\gamma_4}}\right) P((v_2)|EV_0) = \frac{1}{2}\left(\frac{e^{-3.5}+e^{-3}}{e^{-3.5}+e^{-3.5}+e^{-3}+e^{-3}}\right) = 0.25$

Given the initial state $\{0, 0, (v_1, v_2)\}$, the probability of observing path 1 is $P(\sigma_1) + P(\sigma_2)$ and the probability of observing path 2 is $P(\sigma_3) + P(\sigma_4)$.

Table 2 reports the likelihood of observing sequences of states and paths predicted by the two models.

**Table 2** Summary of the results for the illustrative example

|  | Sequence 1 | Sequence 2 | Sequence 3 | Sequence 4 |
|---|---|---|---|---|
| Model | Path 1 | | Path 2 | |
| Recursive model | 0.1345 | 0.25 | 0.3655 | 0.25 |
|  | 0.3845 | | 0.6155 | |
| Non-recursive model | 0.1888 | 0.25 | 0.3112 | 0.25 |
|  | 0.4388 | | 0.5612 | |

We can see from the table that the prediction results of the two models are different. The network is dynamic and stochastic given initial state $\{0, 0, (v_1, v_2)\}$. The recursive logit model assigns more extreme probabilities among the two paths. The difference of the choice probability between the two models is due to the sources of the randomness, i.e., state probability and choice probability, and how they are incorporated in the decision process.

## 4. Comparison of the Two Models

Recursive logit model introduces a random error term associated with each action in the choice set of all outgoing links at each intermediate state. Travelers choose the next link to maximize the sum of instantaneous random utility of the link and expected downstream utility at each intermediate state and expected downstream utilities are identified from Bellman equations. For non-recursive model, routing policy choice happens only once at the origin. In other words, there is already a chosen routing policy for travelers before the trip and the adaptation en route is an execution of the chosen routing policy. This is implemented by first generating the choice set of all reasonable routing policies with some routing policy generation algorithm and then assigning probabilities among routing policies in the choice set by a multinomial logit model (Gao, 2005; Gao et al., 2008).

The three questions we want to address are: (1) What factors lead the difference between the two models? (2) When will the two models predict the same route probabilities? (3) How are the two models different from each other? Are there some regularity of the difference between the two models?

We are guided by intuition that the difference between the two models results from the interplay of two sources of randomness: nature's probability and choice probability. We start our analysis with two benchmark cases for this scenario: (1) deterministic network (2) deterministic choice.

## 4.1. Equivalence of the two models

**Case 1. Deterministic network**

**Proposition 1.** *When the network is deterministic, the two models are equivalent.*

**Proof**. When the network is deterministic, the next state is completely determined by the chosen action. The problem now becomes a fixed path choice problem instead of a routing policy choice problem. It is shown that the recursive model is equivalent to the multinomial logit model at the origin in deterministic network by Fosgerau et al. (2013). **QED.**

**Case 2. Deterministic choice**

**Proposition 2.** *When the scale parameter µ of the error term approaches zero, i.e., the routing policy choice is deterministic, the recursive and the non-recursive models are equivalent.*

**Proof.** The Bellman's equation for the recursive model collapses to the Bellman's equation for the optimal routing policy problem (Gao & Chabini, 2006).

$$V_n^d(k,t,EV) = \max_{a \in A(k)}[\omega_n(a|k,t,EV) + \sum_{EV' \in EV(t')} V_n^d(a,t',EV') P_r(EV'|EV)]\} \quad (12)$$

The solution to this system of equations is the optimal routing policy, which will be assigned with probability 1 by the non-recursive model. This is equivalent to assigning probability 1 to the next link that achieves the maximum in the RHS of the equation at each state, the result of the recursive model. **QED.**

## 4.2. Exploration of the Difference between the Two Models

We now turn our attention to explore how the prediction results of these two models are different from each other. We extend our analysis by allowing the existence of both sources of the randomness.

Consider the choice probability at the end of link 1(node *b*) in Figure 1. There are two possible states at node *b* with two parallel outgoing links. For the sake of convenience, let state 1 denote {1, 1, $v_1$} with a probability of *p* and state 2 denote {1, 1, $v_2$} with a probability of 1-*p*. The key link travel times also play an import role in travelers' route choice behavior. In order to write down the analytical expression of the route choice probability predicted by the two models, we introduce the following parameters:
*a*: travel time on link 2 at time 1 in state 1, $a > 0$;

$b$: travel time on link 2 at time 1 in state 2, $b > 0$;
$a + x$: travel time on link 3 at time 1 in state 1, $x > -a$;
$b + y$: travel time on link 3 at time 1 in state 2, $y > -b$.

The regularity of the difference between the two models is not explicit. However, the numerical results in Table 2 point us a possible direction: either of the two models could predict more extreme probabilities under some specific assumptions. We use the ratio of the choice probabilities of the two routes (with and without given the initial state) as a performance measure in the following.

Table 3 gives the analytical expressions of the two models.

**Table 3** Ratio of the prediction results ($x > -a$, $y > -b$, $0 < p < 1$)

| Model | $\dfrac{P(link2|state1)}{P(link3|state1)}$ | $\dfrac{P(link2|state2)}{P(link3|state2)}$ | $\dfrac{P(link2)}{P(link3)}$ |
|---|---|---|---|
| Recursive model | $e^x$ | $e^y$ | $\dfrac{p(e^y+1)e^x + (1-p)(e^x+1)e^y}{p(e^y+1) + (1-p)(e^x+1)}$ |
| Non-recursive model | $e^{px}$ | $e^{y-py}$ | $\dfrac{p(e^{y-py}+1)e^{px} + (1-p)(e^{px}+1)e^{y-py}}{p(e^{y-py}+1) + (1-p)(e^{px}+1)}$ |

There are three cases as follows:
(1) When $x = 0$ & $y = 0$, the two models predict the same route choice probabilities in any state.
(2) When $x > 0$ & $y > 0$, link 2 is better than link 3 in in any state; **Appendix** tells us that the recursive model predicts more extreme route choice probabilities in any state.
When $-a < x < 0$ & $-b < y < 0$, link 3 is better than link 2 in any state. This is the same with the case when $x > 0$ & $y > 0$ by switching the two routes; the recursive model predicts more extreme route choice probabilities in any state.
(3) When $x > 0$ & $-b < y < 0$, link 2 is better than link 3 at state 1 and link 3 is better than link 2 at state 2; either of the two models could predict more extreme probabilities.
When $-a < x < 0$ & $y > 0$, link 3 is better than link 2 at state 1 and link 2 is better than link 3 at state 2. This is the same with the case when $x > 0$ & $-b < y < 0$ by switching the two routes; either of the two models could predict more extreme probabilities.

To summarize, when a route has state-wise stochastic dominance[b] over the other, the recursive model predicts more extreme route choice probabilities; when the two routes are non-dominated, the relationship can go either way. Note that this result only applies to the network in Figure 1.

---

[b] State-wise dominance (also known as state-by-state dominance) is defined as follows: gamble A is state-wise dominant over gamble B if A gives a better outcome than B in every possible future state.

## 4.3. Discussions of Applications to Real Networks

In this section we provide a discussion of the pros and cons in applying the two routing policy models to real networks, in terms of the computational efficiency of estimation and prediction, and the flexibility in systematic utility specification, and correlation modeling. The non-recursive model is a straightforward extension of the non-recursive path choice model based on choice set generation, and thus also inherits some of the pros and cons of the path model.

### 4.3.1. Computational Efficiency

The non-recursive model requires choice set generation that involves repeated executions of the optimal routing policy algorithm. Once the choice sets are generated and attributes calculated, the estimation can be done with an existing discrete choice model estimation package. In contrast, the recursive model avoids choice set generation, but the estimation requires solving for the value functions for each trial values of the unknown parameters. The optimal routing policy algorithm and the value function solution algorithm usually take comparable amount of time, so the relative efficiency of the two models depends on the number of times these algorithms need to be executed, which conceivably depend on the problem.

The recursive model is superior in computational efficiency in prediction as no choice set generation is needed and it is simply a walk through the network. The non-recursive model requires choice set generation, but if the process is done only once for multiple simulations then it is not a major concern. The computer memory required to store generated routing policies could be a concern in very large networks.

### 4.3.2. Systematic Utility Specification

The non-recursive model can accommodate a wide range of systematic utility specifications, including non-additive attributes. The recursive model however by design can only accommodate additive attributes, which make extra efforts needed to include important attributes such as travel time variability (generally not additive when correlation among link travel times exist) and to employ non-linear utility function such as the prospect theory (Tversky & Kahneman, 1992).

### 4.3.3. Correlation Modeling

Any techniques for modeling correlation in a non-recursive path choice model can be applied to the non-recursive routing policy model, which enables the leverage of the rich literature in this aspect. A nested recursive logit model is recently developed in Mai et al. (2015), however, the study of accounting for correlation in recursive model is still in its early stage and significant research efforts are needed.

## 5. Conclusions and Future Directions

This research studies the routing policy choice problems in an STD network. Two types of routing policy choice models are presented with POI: recursive logit model and non-recursive logit model. We then give a theoretical discussion on how these two modeling approaches can be compared. It is shown that the two models are equivalent in a deterministic network or with deterministic choice. We also use an illustrative example to explore the difference between the two models when both sources of stochasticity exist, and find that when a route has state-wise stochastic dominance over the other, the recursive model predicts more extreme choice probabilities. The relation can go either way when the two routes are non-dominated. Even though we cannot tell which model is more appropriate unless we have real data, our theoretical analysis provides a basis to empirically compare the two models.

The very first next step is the application of the recursive model in a real network. The recursive logit model has the advantage over the traditional sampling approaches that it can be both consistently estimated and used for prediction without generating choice set. This is potentially very useful for e.g. traffic simulation applications where the number of paths that can be stored is restricted by available memory.

Currently we formulate the two modes under POI assumption. However, realistic information situations are generally limited in scope temporally and/or spatially, which is called partial online information. Thus a future direction of interest is to study the problem with partial online information.

Another direction is to extend the model with more realistic descriptions of risk attitudes by applying the cumulative prospect theory (CPT) proposed by Tversky & Kahneman (1992). In this research, we formulate the two models under an assumption of fixed attitude toward risks. However, in a typical risky traffic network risk attitudes play an important role in the decision process.

**Appendix A**

For the sake of convenience, we denote
$P(l_2) = $ probability of choosing link 2 at the end of link1;
$P(l_3) = $ probability of choosing link 2 at the end of link1;
$P(l_2|s_1) = $ probability of choosing link 2 condition on state1 at the end of link1;
$P(l_3|s_1) = $ probability of choosing link 3 condition on state1 at the end of link1;
$P(l_2|s_2) = $ probability of choosing link 2 condition on state2 at the end of link1;

$P(l_3|s_2)$ =probability of choosing link 3 condition on state2 at the end of link1.

Throughout this Appendix, we use a superscript of $r$ denote the probabilities for the recursive model while a superscript of $nr$ denote the probabilities for the non-recursive logit model.

When $x>0$ & $y>0$,

$$\frac{P^r(l_2|s_1)}{P^r(l_3|s_1)} = e^x > 1 \Rightarrow P^r(l_2|s_1) > P^r(l_3|s_1) \tag{1}$$

$$\frac{P^r(l_2|s_2)}{P^r(l_3|s_2)} = e^y > 1 \Rightarrow P^r(l_2|s_2) > P^r(l_3|s_2) \tag{2}$$

$$\frac{P^{nr}(l_2|s_1)}{P^{nr}(l_3|s_1)} = e^{px} > 1 \Rightarrow P^{nr}(l_2|s_1) > P^{nr}(l_3|s_1) \tag{3}$$

$$\frac{P^{nr}(l_2|s_2)}{P^{nr}(l_3|s_2)} = e^{y-py} > 1 \Rightarrow P^{nr}(l_2|s_2) > P^{nr}(l_3|s_2), \tag{4}$$

Thus
$$P^r(l_2) - P^r(l_3) = p[P^r(l_2|s_1) - P^r(l_3|s_1)] + (1-p)[P^r(l_2|s_2) - P^r(l_3|s_2)] > 0$$
$$P^{nr}(l_2) - P^{nr}(l_3) = p[P^{nr}(l_2|s_1) - P^{nr}(l_3|s_1)] + (1-p)[P^{nr}(l_2|s_2) - P^{nr}(l_3|s_2)] > 0$$

Note that

$$\frac{P^r(l_2|s_1)}{P^r(l_3|s_1)} / \frac{P^{nr}(l_2|s_1)}{P^{nr}(l_3|s_1)} = e^x/e^{px} = e^{(1-p)x} > 1 \Rightarrow \frac{P^r(l_2|s_1)}{P^r(l_3|s_1)} > \frac{P^{nr}(l_2|s_1)}{P^{nr}(l_3|s_1)} > 1 \tag{5}$$

$$\frac{P^r(l_2|s_2)}{P^r(l_3|s_2)} / \frac{P^{nr}(l_2|s_2)}{P^{nr}(l_3|s_2)} = e^y/e^{y-py} = e^{py} > 1 \Rightarrow \frac{P^r(l_2|s_2)}{P^r(l_3|s_2)} > \frac{P^{nr}(l_2|s_2)}{P^{nr}(l_3|s_2)} > 1 \tag{6}$$

which imply that the recursive logit model predicts more extreme probabilities at both state 1 and 2. This is equivalent to (

$$\frac{P^r(l_2|s_1)}{P^r(l_3|s_1)} > \frac{P^{nr}(l_2|s_1)}{P^{nr}(l_3|s_1)} \Rightarrow \frac{1-P^r(l_3|s_1)}{P^r(l_3|s_1)} > \frac{1-P^{nr}(l_3|s_1)}{P^{nr}(l_3|s_1)} \Rightarrow \frac{1}{P^r(l_3|s_1)} > \frac{1}{P^{nr}(l_3|s_1)} \Rightarrow$$
$$P^r(l_3|s_1) < P^{nr}(l_3|s_1) \Rightarrow 1 - 2P^r(l_3|s_1) > 1 - 2P^{nr}(l_3|s_1) \ )$$

$$P^r(l_2|s_1) - P^r(l_3|s_1) > P^{nr}(l_2|s_1) - P^{nr}(l_3|s_1) > 0 \tag{7}$$
$$P^r(l_2|s_2) - P^r(l_3|s_2) > P^{nr}(l_2|s_2) - P^{nr}(l_3|s_2) > 0 \tag{8}$$

Next we show that the recursive model would predict more extreme probabilities.

$$[P^r(l_2) - P^r(l_3)] - [P^{nr}(l_2) - P^{nr}(l_3)]$$
$$= \{[p * P^r(l_2|s_1) + (1-p) * P^r(l_2|s_2)] - [p * P^r(l_3|s_1)] + (1-p) * P^r(l_3|s_2)\} -$$
$$[p * P^{nr}(l_2|s_1) + (1-p) * P^{nr}(l_2|s_2)] - [p * P^{nr}(l_3|s_1)] + (1-p) * P^{nr}(l_3|s_2)\}\}$$
$$= p * \{[P^r(l_2|s_1) - P^r(l_3|s_1)] - [P^{nr}(l_2|s_1) - P^{nr}(l_3|s_1)]\} + (1-p) *$$
$$\{[P^r(l_2|s_2) - P^r(l_3|s_2)] - [P^{nr}(l_2|s_2) - P^{nr}(l_3|s_2)]\} \tag{9}$$

With Equation (5), (6) and $0 < p < 1$, one gets

$$[P^r(l_2) - P^r(l_3)] - [P^{nr}(l_2) - P^{nr}(l_3)] > 0 \qquad (10)$$

In addition with Equation (5) and (6)

$$P^r(l_2) - P^r(l_3) > P^{nr}(l_2) - P^{nr}(l_3) > 0 \qquad (11)$$

which implies that the recursive model predicts more extreme probabilities.

**References**

Bellman, R. (1957). *A Markovian decision process* (Tech. Rep.). DTIC Document.

Ding, J., Gao, S., Jenelius, E., Rahmani, M., Huang, H., Ma, L., ... & Ben-Akiva, M. (2014). Routing Policy Choice Set Generation in Stochastic Time-Dependent Networks: Case Studies for Stockholm, Sweden, and Singapore.*Transportation Research Record: Journal of the Transportation Research Board*, (2466), 76-86.

Ding, J., Gao, S., Jenelius, E., Rahmani, M., Pereira, F., & Ben-Akiva, M. (2015). Latent-Class Routing Policy Choice Model with Revealed-Preference Data. In *Transportation Research Board 94th Annual Meeting* (No. 15-1963).

Fosgerau, M., Frejinger, E., & Karlstrom, A. (2013). A link based network route choice model with unrestricted choice set. *Transportation Research Part B: Methodological*, *56*, 70-80.

Frejinger, E. (2008). *Route choice analysis: data, models, algorithms and applications* (Doctoral dissertation, École Polytechnique Federale de Lausanne).

Gao, S. (2005). *Optimal adaptive routing and traffic assignment in stochastic time-dependent networks* (Doctoral dissertation, Massachusetts Institute of Technology).

Gao, S., & Chabini, I. (2006). Optimal routing policy problems in stochastic time-dependent networks. *Transportation Research Part B: Methodological*,*40*(2), 93-122.

Gao, S., Frejinger, E., & Ben-Akiva, M. (2008). Adaptive route choice models in stochastic time-dependent networks. *Transportation Research Record: Journal of the Transportation Research Board*, (2085), 136-143.

Gao, S., & Huang, H. (2009). Is More Information Better for Routing in an Uncertain Network?. In *Transportation Research Board 88th Annual Meeting*(No. 09-1315).

Gao, S., & Huang, H. (2012). Real-time traveler information for optimal adaptive routing in stochastic time-dependent networks. *Transportation Research Part C: Emerging Technologies*, *21*(1), 196-213.

Mai, T., Fosgerau, M., & Frejinger, E. (2015). A nested recursive logit model for route choice analysis. *Transportation Research Part B: Methodological*,*75*, 100-112.

McFadden, D. (1978). *Modelling the choice of residential location* (pp. 75-96). Institute of Transportation Studies, University of California.

Ramming, M. S. (2002). *Network knowledge and route choice* (Doctoral dissertation, Massachusetts Institute of Technology).

Tversky, A., & Kahneman, D. (1992). Advances in prospect theory: Cumulative representation of uncertainty. *Journal of Risk and uncertainty*,*5*(4), 297-323.